\documentclass[twocolumn]{aastex62}
                   
\usepackage{apjfonts}
\usepackage[T1]{fontenc}
\usepackage{color}
\usepackage{amsmath,amstext}
\usepackage{hyperref}
\usepackage{natbib}
\usepackage[normalem]{ulem}

\begin{document}

\title{GIANT PLANET SWAPS DURING CLOSE STELLAR ENCOUNTERS}

\author{Yi-Han Wang}
\affiliation{Department of Physics and Astronomy, Stony Brook University, Stony Brook, NY, 11794, USA}

\author{Rosalba Perna}
\affiliation{Department of Physics and Astronomy, Stony Brook University, Stony Brook, NY, 11794, USA}
\affiliation{Center for Computational Astrophysics, Flatiron Institute, 162 5th Avenue, New York, NY 10010, \
USA}

\author{Nathan W. C. Leigh}
\affiliation{Departamento de Astronom\'ia, Facultad de Ciencias F\'isicas y Matem\'aticas,
Universidad de Concepci\'on, Concepci\'on, Chile}
\affiliation{Department of Astrophysics, American Museum of Natural History, Central Park West and 79th Street, New York, NY 10024}

\begin{abstract}
    The discovery of planetary systems outside of the solar system has
challenged some of the tenets of planetary formation. Among the
difficult-to-explain observations, are systems with a giant planet
orbiting a very-low mass star, such as the recently discovered GJ~3512b
planetary system, where a Jupiter-like planet orbits an $M$-star in a
tight and eccentric orbit.  Systems such as this one are not predicted
by the core accretion theory of planet formation.  Here we suggest a
novel mechanism, in which the giant planet is born around a more
typical Sun-like star ($M_{*,1}$), but is subsequently exchanged
during a dynamical interaction with a flyby low-mass star ($M_{*,2}$).
We perform state-of-the-art $N$-body simulations with
$M_{*,1}=1M_\odot$ and $M_{*,2}=0.1M_\odot$ to study the statistical
outcomes of this interaction, and show that exchanges result in high
eccentricities for the new orbit around the low-mass star, while about
half of the outcomes result in tighter orbits than the planet had
around its birth star.  We numerically compute the cross section for
planet exchange, and show that  an upper limit for the probability per planetary system to have
undergone such an event is  $\Gamma\sim 4.4
(M_{\rm c}/100M_\odot)^{-2}(a_{\rm p}/{\rm AU})  (\sigma/1\,{\rm km}\,{\rm s}^{-1})^{5}$~Gyr$^{-1}$, where $a_{\rm p}$ is the planet semi-major axis around the birth star, 
$\sigma$  the velocity dispersion of the star cluster, and $M_{\rm c}$ the total mass of the star cluster. Hence these planet exchanges could be relatively common for stars born in open clusters and groups, should already be observed in the exoplanet database, and provide new avenues to create unexpected planetary architectures.
\end{abstract}

\section{Introduction}

The discovery of several thousands of exoplanetary systems
has clearly shown that planetary architectures are
considerably more varied than originally thought \citep{Bathala2013}, and
that planet formation models built to explain our
own solar system fall often short of explaining
features and patterns observed in other worlds.

Among the unexpected findings is the recent
discovery of a giant planet orbiting a very-low-mass
star \citep{Morales2019}, the M dwarf GJ 3512b, with a mass of $0.12\,M_\odot$.
The planet, of minimum mass $M_{\rm p}\sin =0.463M_{\rm Jup}$, 
is on an eccentric (eccentricity $e=0.435$) and
tight orbit (semi-major axis $a_J=0.338$~AU).

The discovery of this system is not unique in the current exoplanet set: a giant planet of mass $M_{\rm p}=0.63 M_{\rm Jup}$ 
orbiting a very low mass host (a brown dwarf of mass $M=0.06 M_\odot$) was discovered in the microlensing event MOA-bin-29 \citep{Kondo2019}, and systems of this kind were reported since the early days of planet observations \citep{Delfosse1998}.

Planetary systems like these ones pose a serious challenge to the standard core accretion theory of planet formation \citep{Mizuno1980,
Bodenheimer1986,
Laughlin2004}.
It has in fact been shown \citep{Laughlin2004} that the formation of Jupiter-mass planets orbiting $M$-dwarf stars is highly inhibited at all radial locations, in stark contrast to solar-type stars. 
More recent work \citep{Miguel2019} further confirmed that in the planetesimal accretion scenario a system like GJ~3512b cannot be formed.
\citet{Morales2019} further showed that the pebble accretion theory \citep{Johansen2019} also fails in explaining the configuration of this planetary  system. According to the pebble theory, giant planets accrete upon the formation of a core of at least 5 Earth masses. However, in a system with a low-mass star, migration is high and prevents the core to grow to much large sizes.

A possible explanation of systems like GJ~3512b, if its evolution has
proceeded completely in isolation, involves the onset of the
gravitational instability in the early phases of planet formation,
when the proto-planetary disk was still relatively massive
\citep{Boss2006,Morales2019}. However, for typical values of the disk-viscosity,
fragmentation occurs in the outer parts of the disk, on the order of
tens of AU.  Hence this model also requires substantial migration to
have occurred. High eccentricities are not naturally predicted
via this mechanism.

Here we propose a novel model to explain the properties of GJ~3512,
motivated by the fact that many (if not most) stars are born within
OB associations or in star clusters (e.g. \citealt{
Lada2003}). Even for the solar system, studies of the abundances of isotopes have
led to the suggestion that it used to be part of a star cluster \citep{Adams2001}.  Clusters are
generally thought to dissolve within 20-50~Myr; however, in the absence of external perturbations, they can be long-lived \citep{deGrijs2009}; and in fact, long-lived clusters are known
to exist (e.g.  the Hyades and Presepe are about 600 Myr old, NGC 6811 is about 1 Gyr, NGC 6819 is about 2.5 Gyr, etc.;  \citealt{Meibom2015,meibom18}).

A number of studies (e.g.  \citealt{Heggie1996,Laughlin1998, Davies2001,Bonnell2001,Thies2005,Fregeau2006,Olczak2010,Chatterjee2012,Portegies2015,
Cai2017,Cai2018,Rice2018,Cai2019,
vanelteren2019, Flammini2019}) have shown how the evolution of planetary systems in
interacting environments may  provide alternative
formation paths for planet properties which are difficult to account for
by current theories of planetary formation.  For example, internal dynamical interactions in multi planet systems may have played a role in producing eccentric planetary orbits (e.g.\citealt{Rasio1996,Weidenschilling96,DelaFuente1997,Chatterjee2008, Juric2008, Beauge2012}), in altering the distribution of mutual inclinations \citep{Chatterjee2008,Boley2012}, in shrinking the orbits of giants leading to hot Jupiters (e.g.\citealt{Nagasawa08, Shara2016, Hamers2017}), or in creating free-floating planets (e.g. \citealt{Chatterjee2008, Juric2008}) which, upon capture, reside on very wide orbits \citep{Perets2012}.

Here we suggest a novel dynamical explanation to create a system such
as GJ~3512b: the giant planet was originally born around a typical,
Sun-like ($G$) star. However, during its lifetime, a flyby by a
low-mass ($M$) star resulted in the planet being swapped between the
two stars. 
Indeed, given the relatively high abundance of low-mass stars compared to solar-type ones \citep{Scalo1979}, a scenario in which a 'solar-type' planetary system is perturbed by a flyby of a low-mass star is the most common one to happen. We note that a planet exchange from a a main sequence star to a neutron star-white dwarf binary system was suggested by \citet{Fregeau2006} to explain the planetary system PSR~B1620-26.

We perform highly-accurate $N$-body simulations to study
the frequency of this planet exchange from the $G$ to the $M$ star, as
well as the properties of the resulting planet and planet+star system.
We find that, for stars born in associations, the rate of this special
dynamical interaction is consistent with a handful of systems in the
current exoplanet set. The high eccentricity is naturally explained
via this mechanism, and tighter orbits than what the planet had
around its birth star are found in about half of the exchanges.

Our paper is organized as follows: Sec~2 describes the computation of the cross-section for the planet exchange, 
and hence the rate for this dynamical mechanism. We hence present (Sec.~3) the 
properties of the planetary system formed after the swap, and discuss them in the context of GJ~3512b. We summarize and conclude in Sec~4.

\begin{figure}
 \includegraphics[width=1\columnwidth]{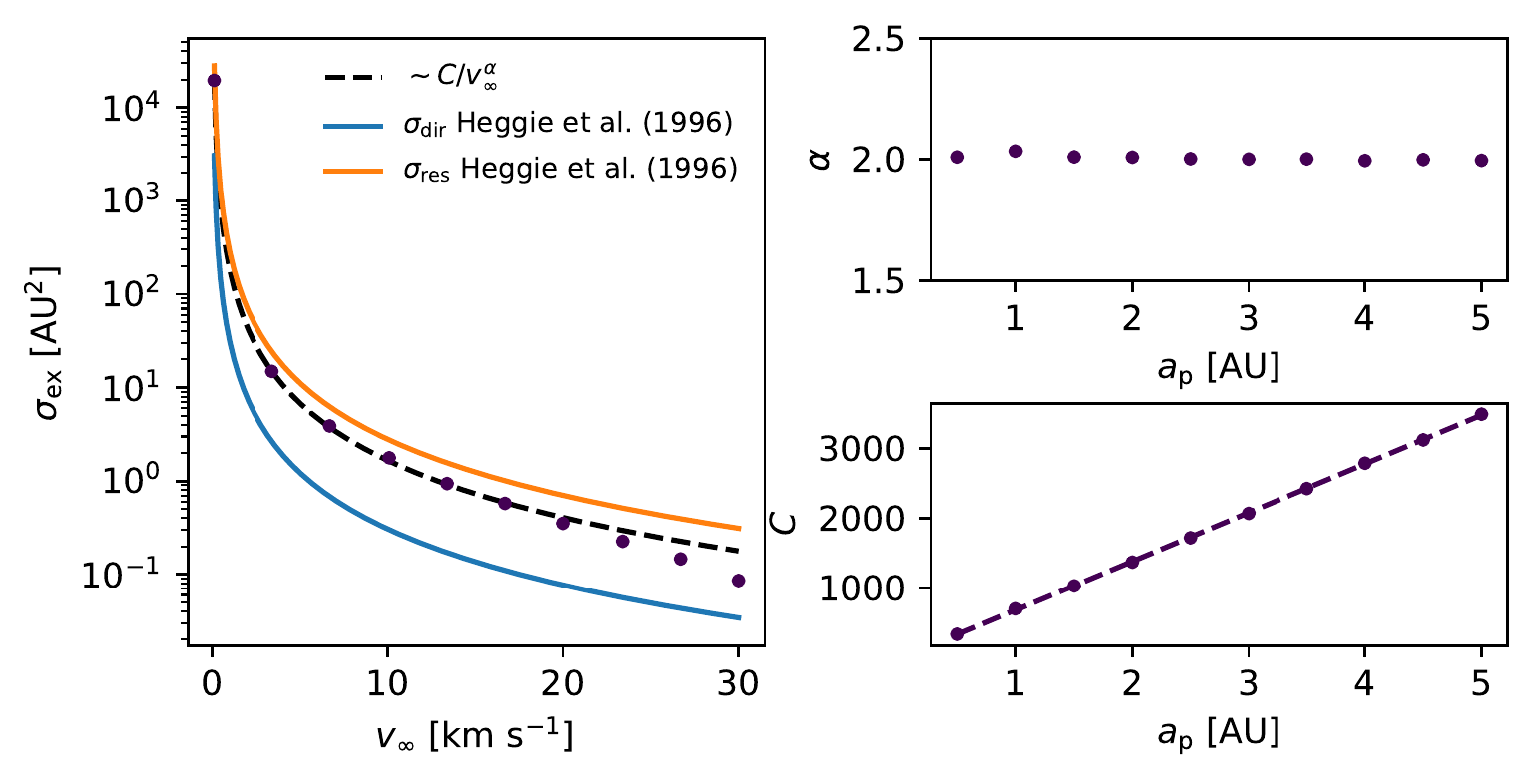}
  \caption{Results from numerical experiments and their best fit. Left: Cross section for exchange of a giant planet from an $1M_\odot$ star to a $0.1M_\odot$ one, for an initial orbital separation $a_{\rm p}=1$~AU,   which falls in the regime $v_{\rm c}<v_\infty<v_{\rm orb}$.
   As a reference, a comparison is made with the analytical formulas provided by \citet{Heggie96} in the regime $v_\infty<v_{\rm c}$.
  Right: Dependence of the cross-section for exchange on the initial planet separation $a_{\rm p}$. } 
 \label{fig:fitting}
\end{figure}

\section{Cross section for planet exchange}

The rate for planet exchange is given by
\begin{equation}
\Gamma_{\rm ex} \sim \sigma_{\rm ex} n_* \bar{v}\,,
\end{equation}
where $\sigma_{\rm ex}$ is the cross-section for this mechanism, $n_*$ is the number density of stars in the environment under consideration, and $\bar{v}$ is the typical mean relative velocity in that environment.

In order to compute the cross section, we perform 
numerical scattering experiments with the very high precision few-body code {\tt SpaceHub} (details in \citealt{Wang2018,Wang2019}). The code implements cutting edge chain-regularization \citep{Mikkola1993} and positive round-off error compensation in order to treat the high mass ratio of the star-planet systems that with traditional integrators often result in inaccurate results.

The cross section is calculated as a function of $M_{*,1}$, $M_{*,2}$, $M_{\rm p}$, $a_{\rm p}$ and $V_\infty$, where $M_{*,1}$ is the mass of the $G$-type star that initially hosts the planet, $M_{*,2}$ is the mass of the $M$-type star that dynamically interacts with the $G$ star, $M_{\rm p}$ is the mass of the planet whose original semi-major axis is $a_{\rm p}$, and $v_\infty$ is the relative velocity at infinity (prior to the scattering) between the centre of mass of the $G$-type star-planet system and the $M$ star.

\begin{figure}
 \includegraphics[width=1\columnwidth]{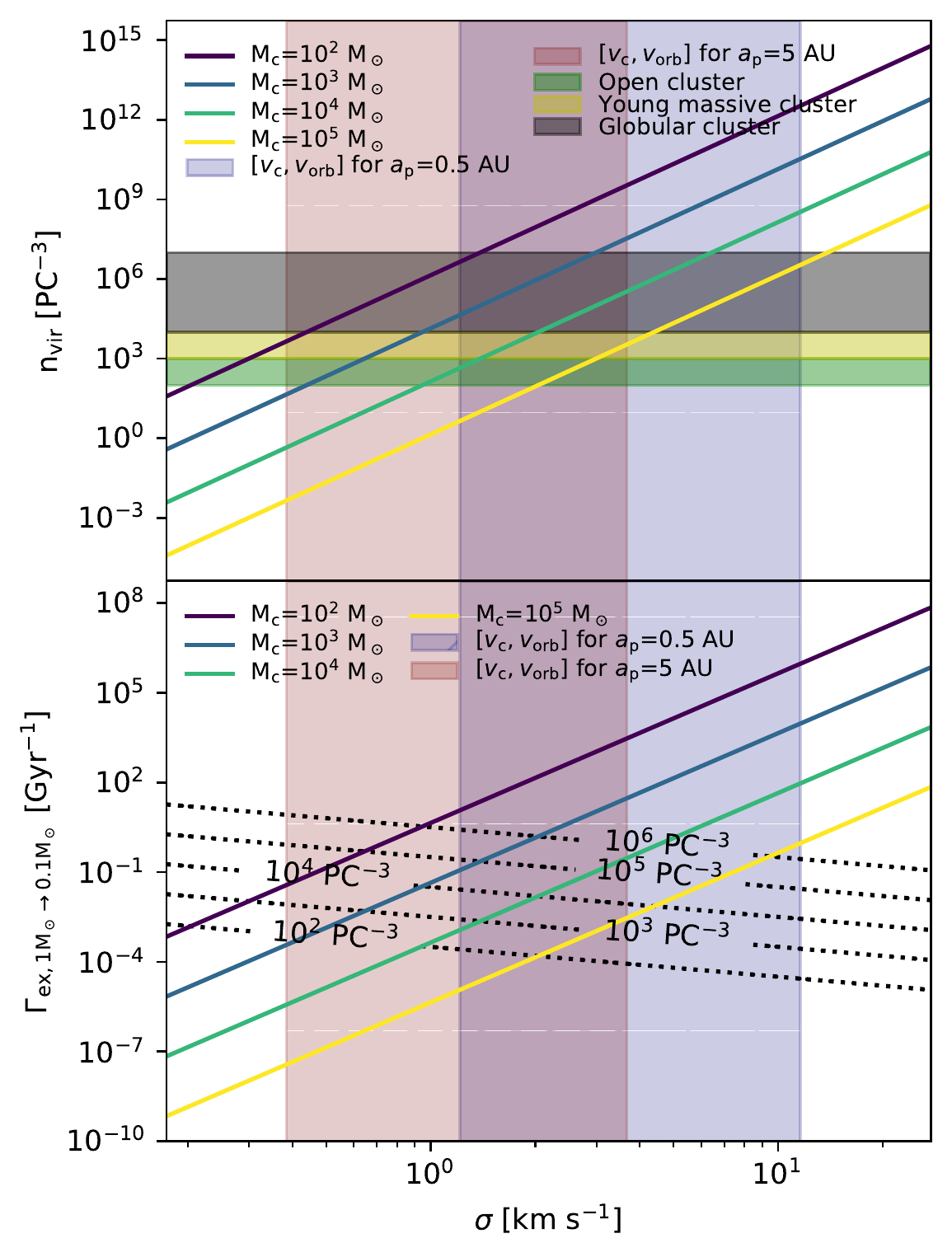}
  \caption{\textit{Upper panel:} Number density of virialized clusters as a function of $\sigma$. \textit{Bottom panel:} Rate of planet exchanges for virialized clusters,  using the cross section in Fig.\ref{fig:fitting}. The dashed lines show the corresponding number densities.}
 \label{fig:rate}
\end{figure}

For each set of values for $M_{*,1}$, $M_{*,2}$, $M_{\rm p}$, $a_{\rm p}$ and $v_\infty$, we perform 
one million scattering experiments between the $G$- star/planet system and the flyby $M$-star. The initial phase parameters of the planet orbit are isothermally distributed, i.e. $\cos(i)$ 
($i$ is the orbital inclination)
is uniformly distributed within $[-1,1]$, while $\Omega$ (longitude of the ascending node), $\omega$ (argument of periapsis) and 
 the mean anomaly $\cal{M}$ 
are all uniformly distributed within $[-\pi, \pi]$.
 The impact parameter $b$ is randomly generated from a distribution uniform in
 $b^2$  within the range $[0, b_{\rm max}]$. The maximum value $b_{\rm max}$, for each combination of  $M_{*,1}$, $M_{*,2}$, $M_{\rm p}$, $a_{\rm p}$ and $v_\infty$, is numerically pre-determined to ensure that all the impact parameters $b<b_{\rm max}$ that may lead to planet exchange are included in the scattering experiment.  
\begin{figure*}
\includegraphics[width=2\columnwidth]{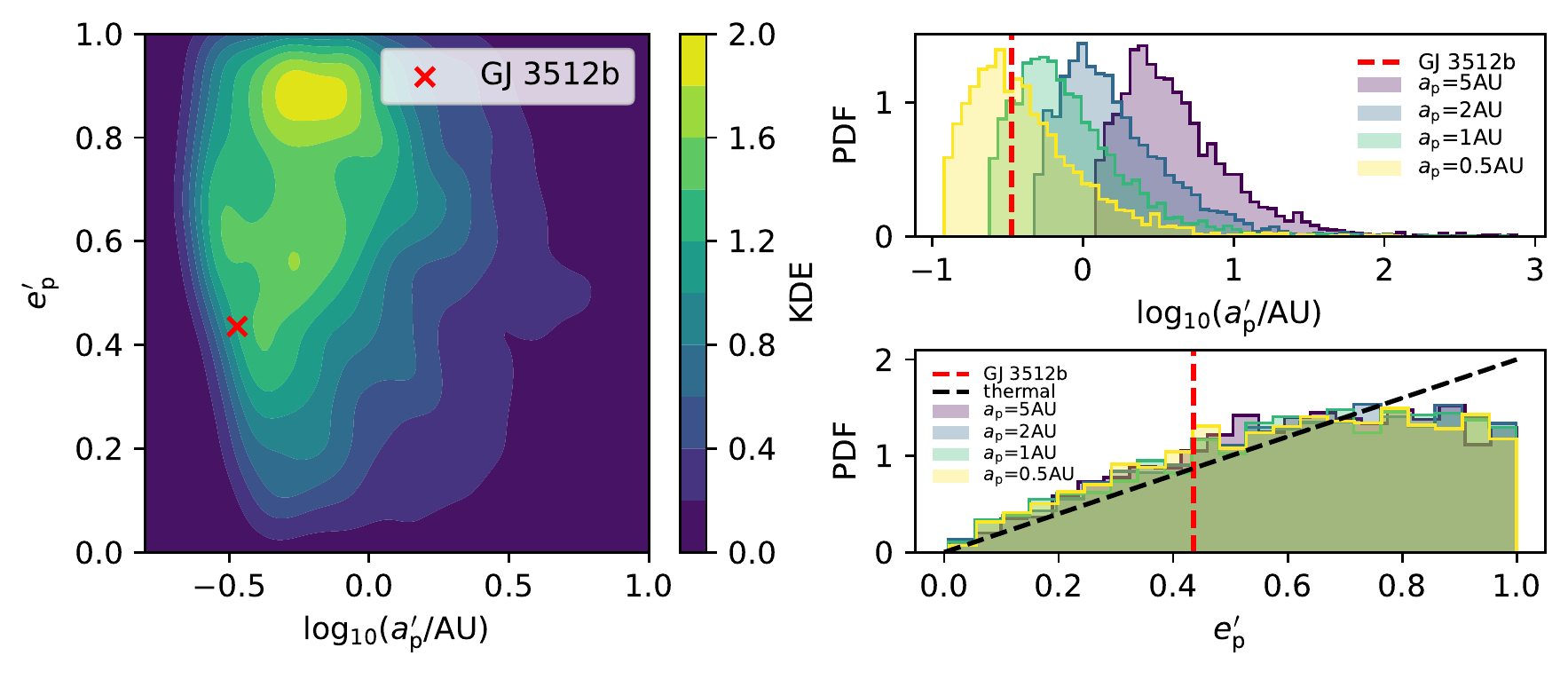}\\
\includegraphics[width=2\columnwidth]{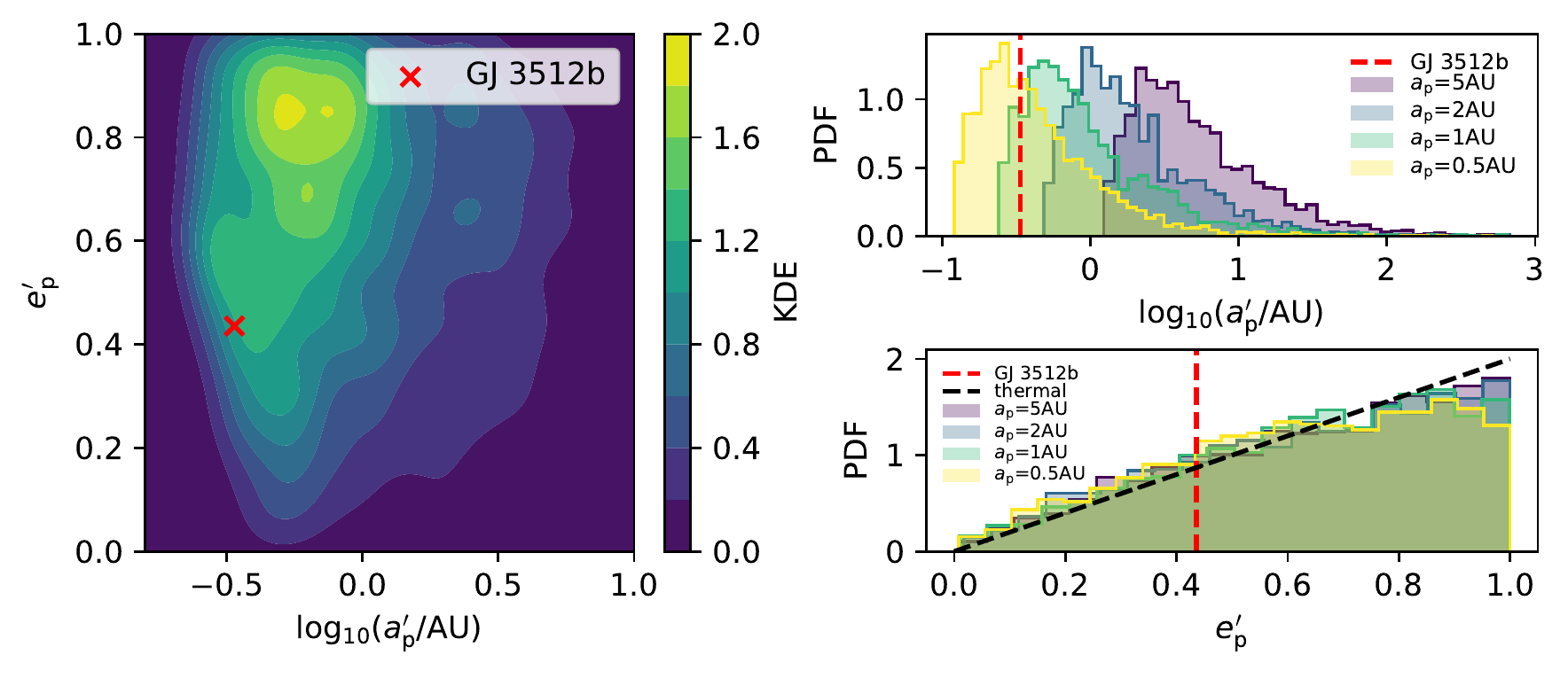}\\
 \includegraphics[width=2\columnwidth]{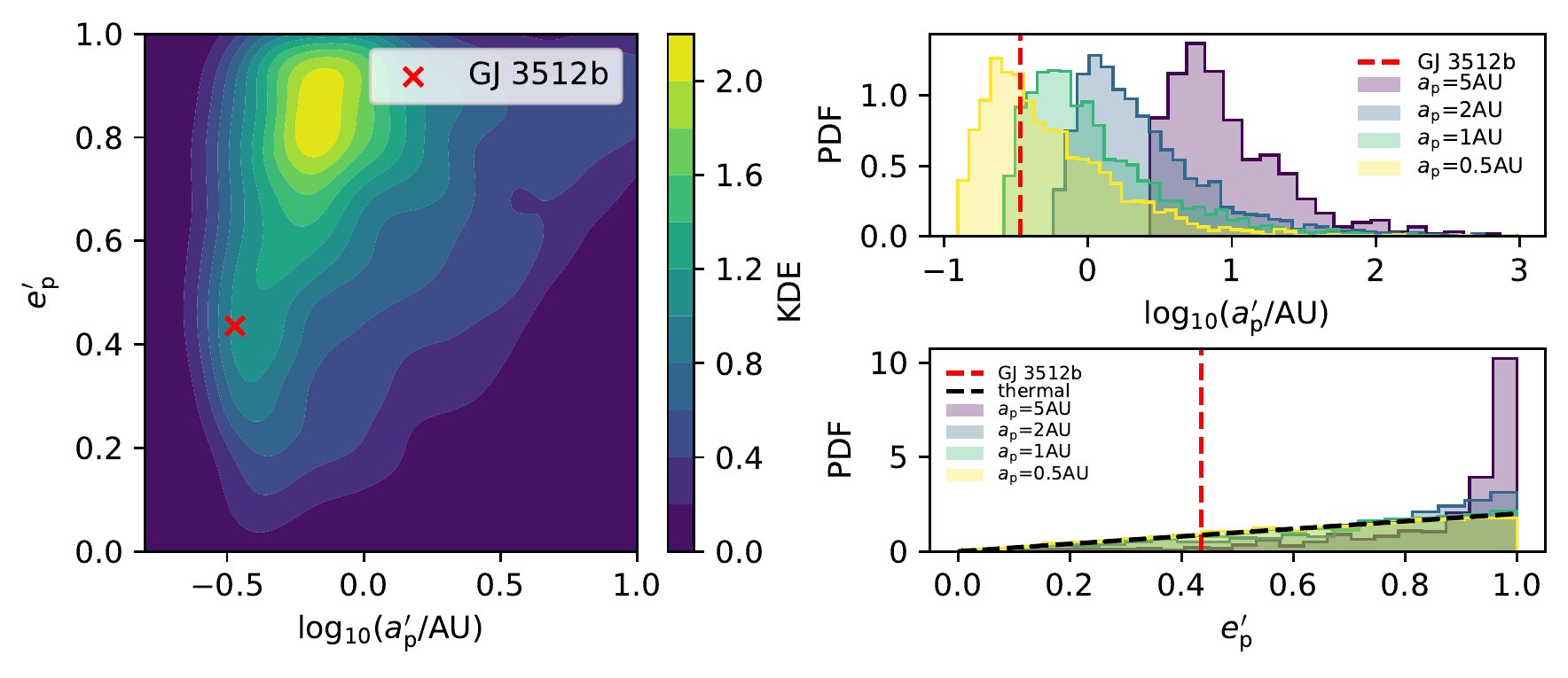}
  \caption{The post scattering orbital separation and eccentricity of the planet after the swap onto the low-mass star, for three values of the relative velocity at infinity: $v_\infty=0.1$~km~s$^{-1}$ (top panel), $v_\infty=3.4$~km~s$^{-1}$ (middle panel) and $v_\infty={13.4}$~km~s$^{-1}$ (bottom panel). Left: 2D probability distribution function (PDF) for the case $a_{\rm p}=1$~AU and $e_{\rm p}=0$. Right: the PDF of the orbital separation (top) and the eccentricity (bottom) for a range of $a_{\rm p}$ with $e_{\rm p}=0$. The observed parameters of GJ~3512b are also shown for reference. }
 \label{fig:orbits}
\end{figure*}

If $N_{\rm tot}$ is the total number of scattering experiments, and $N_{\rm ex}$ the number of outcomes found to be planet swaps, then the cross section for this mechanism is \citep[e.g.][]{hut83}

\begin{equation}
\sigma_{\rm ex} = \pi b_{\rm max}^2\frac{N_{\rm ex}}{N_{\rm tot}}\,,
\end{equation}
with statistical error
\begin{equation}
\Delta\sigma_{\rm ex} = \pi b_{\rm max}^2\frac{\sqrt{N_{\rm ex}}}{N_{\rm tot}}\,.
\end{equation}

In the following, in order to investigate planet swaps as a mechanism to explain systems 
such as GJ~3512, we specialize 
our simulations to the following values:  $M_{*,1}=1M_\odot$, $M_{*,1}=0.1M_\odot$ and $M_{\rm p}=M_{\rm Jup}$, and explore the dependence on $a_{\rm p}$ (original orbital separation around $M_{*,1}$) and $v_\infty$, since these are not directly measured variables.

The left panel of Fig.~\ref{fig:fitting} shows the cross-section calculated from the scattering experiments as a function of $v_\infty$ and for $a_{\rm p}=1$~AU.  We parameterize our fit to these data as $\sigma_{\rm ex} =$ Cv$^{\alpha}$, where C and $\alpha$ are fitting parameters.  The best fit powerlaw to the numerical data is with $\alpha=-2.03$. The numerically-derived cross section falls into the region between the cross section of direct planet exchange $\sigma_{\rm dir}$ in Eq.(13) of \citet{Heggie96}, and the cross section of resonance planet exchange $\sigma_{\rm res}$ in Eq.(15) of \citet{Heggie96}. In the \citet{Heggie96} paper, the cross sections are estimated in the regime $v_\infty < v_{\rm c}$, where $v_c$ is the critical velocity, i.e. the velocity for which the total energy of the system is zero. However, most of the parameter space covered here is in the intermediate regime ($v_{\rm c} < v_\infty < v_{\rm orb}$), where $v_{\rm orb}$ is the initial orbital velocity of the planet. Thus,  their regime does not directly apply to our calculations. This is also evident by the different behavior of the cross sections, which is shown for comparison in Fig.~\ref{fig:fitting}. On the other hand, \citet{Fregeau2006} discussed the cross section of planet exchange in our regime. However, the cross section they provide for planet exchange is for the equal star mass case, unlike  the low star mass ratio that we study here. The right panel of Fig.~\ref{fig:fitting} shows the linear relationship between $a_{\rm p}$ and $\sigma_{\rm ex}$ from the scattering experiments with different values of $a_{\rm p}$. 

For virialized clusters,
\begin{equation}
M_{\rm c}\sim \frac{2R_{\rm c}v^2}{G}\,,
\end{equation}
where $M_{\rm c}$ and $R_{\rm c}$ are the mass and radius of the cluster, respectively, and $v$ is its root mean square velocity. The number density of a virialized cluster can be then estimated as
\begin{equation}
n_{\rm vir}\sim\frac{M_{\rm c}/\bar{m}}{4\pi R_{\rm c}^3/3}\sim\frac{6v^6}{\pi G^3M_{\rm c}^2\bar{m}}\,,
\end{equation}
where $\bar{m}$ is the mean stellar mass in the cluster. The upper panel of Figure~\ref{fig:rate} shows the number density of the virialized cluster with different cluster masses as a function of cluster velocity dispersion $\sigma$. The relationship between $v_\infty$ and $\sigma$ for a Maxwellian-Boltzmann distribution is
\begin{equation}
\sigma=\sqrt{\frac{3\pi-8}{3\pi}}v,\,\,\,\langle v_\infty^2\rangle =\langle 2v^2 \rangle\,.
\end{equation}
Hence, the exchange rate per planetary system can be estimated as
\begin{equation}
\Gamma_{\rm ex} \sim n_{\rm vir}\sigma_{\rm ex} v \sim \frac{6\sigma_{\rm ex}v^7}{\pi G^3M_{\rm c}^2\bar{m}}\,.
\end{equation}
Making use of the fitted $\sigma_{\rm ex}$, and for a typical $\bar{m}=0.5$~$M_\odot$, we  obtain
\begin{equation}
\Gamma_{\rm ex,1M_\odot\rightarrow 0.1M_\odot}\sim 4.4\bigg(\frac{a_{\rm p}}{\rm AU} \bigg)\bigg(\frac{\sigma}{\rm km~s^{-1}}\bigg)^5 \bigg( \frac{M_{\rm c}}{10^2M_\odot}\bigg)^{-2}{\rm Gyr^{-1}}\,.
\end{equation}
The lower panel of Figure~\ref{fig:rate} shows the exchange rate (per planetary system) for $a_{\rm p}=1$~AU and $\bar{m}=0.5 M_\odot$ with different cluster masses. The dashed lines show the corresponding virialized number density, while the color regions indicate the interaction regime $v_{\rm c}<v_\infty<v_{\rm orb}$.

We need to point out that these rates 
should be considered as upper limits. Firstly, the scattering experiments are made with a $1\,M_\odot$ star interacting with a $0.1\,M_\odot$ star.
While these are indeed very common, the interacting stars have a mass distribution, upon which the cross section depends.
Second, due to mass segregation, the more massive stars in a cluster will tend to sink towards the center, while the lighter $G$ and $M$ stars will populate the less dense regions of the cluster. Numerical simulations by \citet{Chatterjee2012} suggest that, due to primarily this reason, about 10\% of all planetary systems around low-mass stars take part in a strong encounter in clusters similar to the open cluster NGC 6791.

Third (and most importantly), 
other dynamical processes, such as planet ejections, compete with planet exchanges during close encounters, and this is a sensitive function of environment.  
A full study of the relative rates of the various processes is deferred to follow up work (Wang et al. 2020, in prep).

In the following, we will present the results
of numerical experiments with three choices of the initial relative velocity: $v_\infty =0.1$~km~s$^{-1}$(for all $a_{\rm p}$, in the hard binary regime where $v_\infty < v_{\rm c}$),
$v_\infty =3.4$~km~s$^{-1}$(for all $a_{\rm p}$ in the intermediate regime where $v_{\rm c} < v_\infty < v_{\rm orb}$)
and $v_\infty =13.4$~km~s$^{-1}$(for $a_{\rm p}=5$ AU, in the soft binary regime where  $v_\infty > v_{\rm orb}(5~{\rm AU})$).
These fully bracket the typical values
of stars born in dense groups \citep{Binney1987,Adams2001}.

\section{Post-scattering properties of the $M$-star/planet system}

The orbital properties of the planet after being exchanged from the $G$ to the $M$-star are displayed in Fig.~\ref{fig:orbits}. The left panel shows the 2D kernel density distribution of the post-scattered semi-major axis $a_{\rm p}^\prime$ and eccentricity $e_{\rm p}^\prime$ of the  planet after the exchange, for a representative case with $a_{\rm p}=1AU$ and $e_{\rm p}=0$. 
Along with the semi-major axis variation, 
the figure shows the
intrinsic high eccentricity produced from the dynamical interaction. During a planet swap, the planet can be transferred from the original $G$-star to the new $M$-star in both prograde ($M$-star flies by in the same direction of the planet orbit) and retrograde ($M$-star flies by in the opposite direction to the planet orbit) orbits. 

The right panels of Fig,\ref{fig:orbits} show the collapsed (1D) probability distribution functions for $a_{\rm p}^\prime$ (top panel) 
and $e_{\rm p}^\prime$ (bottom panel) for a range of initial orbital separations $a_{\rm p}$ of the planet, with $e_{\rm p}=0$ in all cases. 
 In the hard binary regime ($v_\infty=$~0.1~km~s$^{-1}$) where $v_\infty < v_{\rm c}$,  $a_{\rm p}^\prime$, on average, shifts towards the lower end as expected due to hardening. The shapes of the distributions for different $a_{\rm p}$ are almost identical.  In this regime, $e_{\rm p}^\prime$ is distributed more towards lower values compared with the thermal distribution. For different $a_{\rm p}$, the distribution of $e_{\rm p}$ also looks similar.

 To better interpret our results, we note that
\citet{Hills1989} and \citet{Hills1990} (see also \citealt{Fregeau2006})
found  that, for extreme unequal mass scatterings, the hard/soft boundary is more accurately defined by $v_{\rm orb}$ rather than by $v_{\rm c}$. Hence, for intermediate regimes ($v_\infty=$~3.4~km~s$^{-1}$) where $v_{\rm c} < v_\infty < v_{\rm orb}$,  $a_{\rm p}^\prime$, on average, also becomes tighter. In this regime, the distribution of $e^\prime_{\rm p}$ is almost thermal. For $a_{\rm p}=0.5$, 3.4~km~s$^{-1}$ is very close to its $v_{\rm c}\sim$~4.4~km~s$^{-1}$. Thus, the  $e^\prime_{\rm p}$ distribution displays the trend of shifting to the hard regime. 

The case in the bottom panel of Fig.~\ref{fig:orbits}, where $a_{\rm p}=$5 AU and $v_\infty=$~13.4~km~s$^{-1}$, is in the soft binary regime where $v_\infty > v_{\rm orb}$. In this regime, we can see that $a_{\rm p}^\prime$, on average, shifts outwards and the $e_{\rm p}^\prime$ is distributed more towards high values compared to the thermal case.

\begin{figure}
  \includegraphics[width=0.9\columnwidth]{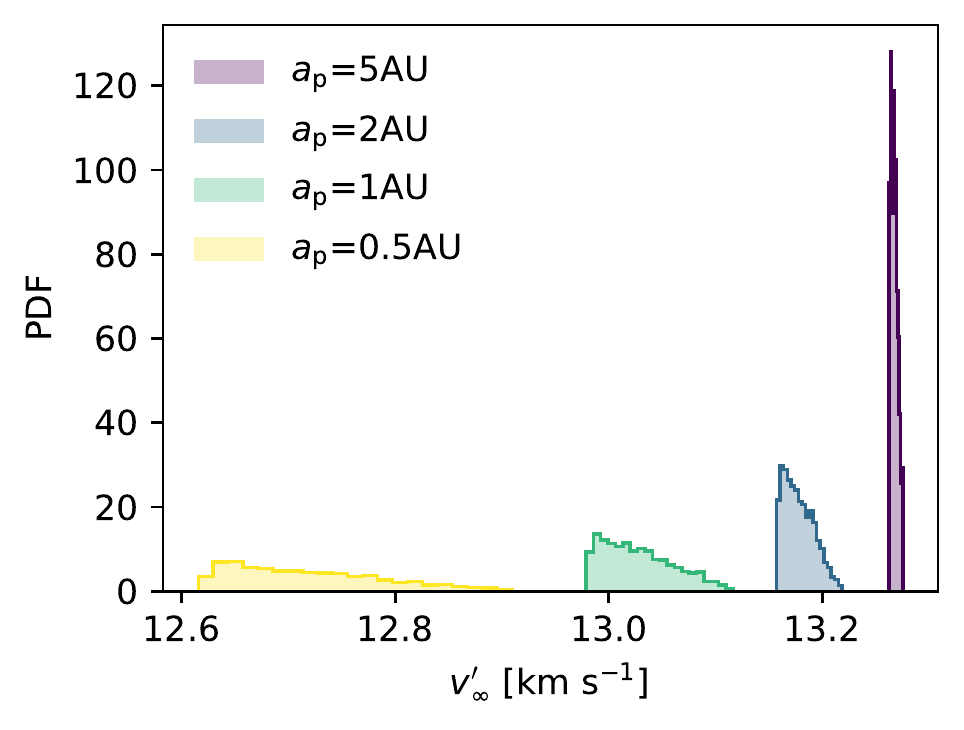}
  \caption{Post-scattering relative velocity of the $M$ and  $G$ stars after the planet swap from the  $G$ to $M$ star, for the case 
   $v_\infty=13.4$~km~s$^{-1}$.}
 \label{fig:velocity}
\end{figure}

The post-scattering relative velocity distribution of the two stars,  after the $M$ star has  acquired the planet from the $G$ star during the dynamical
interaction, is shown in Fig.~\ref{fig:velocity}
for the case $v_\infty=13.4$~km~s$^{-1}$. For this high relative velocity at infinity, the post-scattering relative velocity remains of the same order of magnitude as the initial one, showing only a slight decrease for tighter
captures.  The slight decrease comes from the binding energy shift. The average energy shift can be expressed as
\begin{equation}\label{eq:shift}
\langle \Delta E \rangle \propto \frac{-M_{*2}}{\langle a_{\rm p}^\prime \rangle} - \frac{-M_{*1}}{\langle a_{\rm p} \rangle }\propto \frac{M_{*1}-AM_{*2}}{\langle a_{\rm p} \rangle}\,,
\end{equation}
where $\langle a_{\rm p} \rangle /\langle a_{\rm p}^\prime \rangle =A$ indicates the average semi-major axis shift with $A$ almost identical for $a_{\rm p}=0.5,1$ and $2$ AU. The shifted binding energy will boost/decelerate the centre of mass velocity of the new $M$-star+planet system. For the specific setup studied here, as shown in the bottom right panel of Figure.~\ref{fig:orbits}, we have $1<A<10$. With $M_{*1}=1M_\odot$ and $M_{*2}=0.1M_\odot$, this yields $\langle \Delta E\rangle > 0$. Due to  energy conservation, the kinetic energy of the center of mass of the new star-planet system will decrease, which results in a reduction of $v_\infty$. As clearly shown in Eq.~(\ref{eq:shift}), smaller $a_{\rm p}$ values yield larger energy shifts and a wider dispersion for the same values of 
$M_{*,1}, M_{*,2}$ and
 $A$. 

For the two cases with smaller velocities  
($v_\infty=0.1$~km~s$^{-1}$ and $v_\infty=3.4$~km~s$^{-1}$), the change in binding energy of the planet as it is swapped from the $1\,M_\odot$ star to the $0.1\,M_\odot$ flyby becomes
comparable to or larger than the available kinetic energy in the system, and the two stars remain weakly bound (which is why we do not show their post-scattering relative velocity here). While we do not follow the long-term fate of these weakly bound stars (we are interested in the fate of the planet here), we note that in dense environments these binaries are likely to be eventually disrupted.

For a direct comparison with observations, we note that the velocity of the low-mass star is reflective of the post-scattering
velocity only for a relatively short time. After the host cluster dissolution, the captured planetary system will end up orbiting as an isolated object within the host Galaxy potential.  Hence, the observed velocity of the star will become on the order of the orbital velocity of its original host cluster, imposing only very high relative velocity encounters with other isolated stars.  Thus, the host cluster environment is, in our scenario, needed to ensure low relative velocity interactions, drastically increasing the capture probability per interaction.

\section{Summary}
Motivated by the discovery of planetary systems with gas giants orbiting low-mass stars, which are not explained 
by standard planet formation theories, here we have proposed a novel scenario of a dynamical origin:  The giant is born around a more standard Sun-like star, but gets then captured by a low-mass star during a close encounter. 

We have quantified the occurrence rate of these events, and the statistical properties of the post-scattered systems, via 
highly accurate direct $N$-body simulations, which yielded the (velocity-dependent) cross-section for planet exchange.
For small clusters with total mass $M_{\rm c}\sim10^2-10^3$~$M_\odot$ and
velocity dispersion $1$~km~s$^{-1}$
as typical of star clusters \citep{Adams2001},
exchange rates can be as high as $\sim 0.044-4.4$~Gyr$^{-1}$ per planetary system for a given planet-hosting Sun-like system and an interloper low-mass star, making this mechanism potentially relatively common for stars born in groups.

We find that, after the exchange, the
distribution of planet eccentricity is weighed towards high values, whereas the
  orbital separation  correlates with the initial one that the planet had around its host star, but the distribution is broad.

  Our planet swap mechanism hence provides an alternative path to the formation of gas giants around very-low-mass stars, and naturally predicts some of the observed properties, such as the high orbital eccentricity.

\section*{Acknowledgments}

 We thank the referee for a very thoughtful
and constructive report.
N.~W.~C.~L. acknowledges the generous support of Fondecyt Iniciacion Grant \#11180005.

\bibliographystyle{aasjournal}
\bibliography{biblio.bib}

\end{document}